\documentclass[
 amsmath,amssymb,
 aps,bm
]{revtex4-2}
\usepackage{amsfonts}
\usepackage[caption = false]{subfig}
\usepackage{graphicx}
\usepackage{dcolumn}
\usepackage{bm}
\usepackage{hyperref}
\usepackage{cleveref}
\crefname{figure}{Fig.}{Figs.}
\Crefname{figure}{Fig.}{Figs.}

\def\ltap{\ \raise.3ex\hbox{$<$\kern-.75em\lower1ex\hbox{$\sim$}}\ }
\def\gtap{\ \raise.3ex\hbox{$>$\kern-.75em\lower1ex\hbox{$\sim$}}\ }
\def\lsim{\ \raise.3ex\hbox{$<$\kern-.75em\lower1ex\hbox{$\sim$}}\ }
\def\gsim{\ \raise.3ex\hbox{$>$\kern-.75em\lower1ex\hbox{$\sim$}}\ }

\newcommand{\ie}{{\it i.e.}}
\newcommand{\etc}{{\it etc}}
\newcommand{\be}{\begin{equation}}
\newcommand{\ee}{\end{equation}}
\newcommand{\beq}{\begin{equation}}
\newcommand{\eeq}{\end{equation}}
\newcommand{\bea}{\begin{eqnarray}}
\newcommand{\eea}{\end{eqnarray}}
\newcommand{\bear}{\begin{eqnarray}}
\newcommand{\eear}{\end{eqnarray}}
\newcommand{\mstar}{\ensuremath{M_*}}
\newcommand{\rstar}{\ensuremath{R_*}}
\newcommand{\vecb}[1]{\ensuremath{\mathbf{#1}}}
\newcommand{\eref}[1]{(\ref{#1})}
\newcommand{\gagamgam}{g_{a\gamma\gamma}}

\def\GeV{\,{\rm GeV}}

\def\eV{\,{\rm eV}}
\def\apj{ApJ}

\usepackage[normalem]{ulem}
\usepackage[usenames,dvipsnames]{xcolor}

\begin{document}

\preprint{FERMILAB-PUB-25-0567-T}
\noindent\makebox[\textwidth][r]{\small FERMILAB-PUB-25-0567-T}

\title{
Radio Killed the Axion Star: Constraining Axion Properties with Radio Telescopes
}

\author{Patrick J. Fox$^1$}
\author{Neal Weiner$^2$}
\author{Huangyu Xiao$^{1,3}$}
 \affiliation{$^1$Theoretical Physics Department, Fermilab, Batavia, IL 60510, USA}
\affiliation{$^2$Center for Cosmology and Particle Physics, Department of Physics, New York University, New York, NY 10003, USA}%
\affiliation{$^3$Kavli Institute for Cosmological Physics, University of Chicago, Chicago, IL 60637}

\begin{abstract}
Axion dark matter or any ultralight bosonic dark matter can go through Bose-Einstein condensation due to the large phase density, leading to the formation of axion stars or solitons in dark matter halo centers. The formation rate is enhanced in the presence of the substructures expected in the post-inflationary scenario for the QCD axion or axion-like particles. An axion star will continue to grow until a critical mass is reached, after which it collapses and then explodes, with the emission of relativistic axions, in a process called an ``axinovae.''  There can also be accompanying photon emission due to the stimulated decay of axions in the coherent compact axion star. In axion models with a modest enhancement ($\kappa\sim \mathcal{O}(10)$) of the axion-photon coupling $g_{a\gamma}= \kappa \alpha/(2\pi f_a)$ axinovae will contain a significant flux of radio photons.  We determine the range of parameters over which  axinovae can be detectable with radio transient searches.
\end{abstract}
\maketitle
\section{Introduction}
Since the dawn of history, humans have looked up to the sky and pondered the nature of the luminous objects and what governs their motion. In modern times, we have come to understand the evolution of stars, and the galaxies in which they reside.  The formation of galaxies is dominated by the gravitational potential of dark matter, whose nature remains elusive. In considering scenarios for dark matter, we have become aware of a multitude of possible phenomenological signatures, both subtle and striking.

Light scalar fields appear ubiquitously in field theories, arising as a result of spontaneously broken (approximate) global symmetries. These ``axion'' fields can play important roles in particle physics, not least of which is an elegant solution to the strong CP problem \cite{Weinberg:1977ma,Wilczek:1977pj,PhysRevLett.43.103,Abbott:1982af, Dine:1982ah,Preskill:1982cy, Peccei:2006as}. If these symmetries are broken after inflation, the density field naturally contains $\mathcal{O}(1)$ isocurvature perturbations on small scales corresponding to the horizon size when the axion fields start to oscillate. These can seed dense minihalos, whose cores can condense into dense ``stars'' of the scalar condensate. These stars can evolve as they accrete material from the dense environment, growing until they reach a critical mass, as which point they collapse and explode, producing semi-relativistic axions. This processing of material can lead to important cosmological signals that are similar to decaying dark matter, constraining broad swaths of parameter space \cite{Fox:2023aat}, simply from the conversion of mass to relativistic energy.  Notably, this process relies only on the axion's gravitational- and self-interactions.

In many models, there are couplings beyond just self couplings. If any heavy fields coupling to the axion carry SM charges, it is natural for the scalar to couple via higher dimension operators to photons, gluons and other light fields. An important question is whether these same collapse processes could produce visible SM radiation in the presence of such couplings.
In the presence of the CMB, we shall see that with sufficient coupling, these ``axinovae'' can become interesting sources of electromagnetic radiation, providing new constraints and - possibly - new signals. We call such sources \textit{visible axinovae}, which emit electromagnetic radiation from axion star explosions. Visible axinovae provide more detectable signals than the ordinary axinovae that only emit relativistic axions, expanding the model parameters one can probe with various observations.

The layout of this paper is as follows: in section \ref{sec:formation} we review the formation and collapse of axion stars, which will yield our constraints. In section \ref{sec:EM_emission} we will study the process of EM emission from axinovae.  We compile these in section \ref{sec:observation}, and review what limits present and future observations can place on these scenarios. Finally, in section \ref{sec:conclusion} we conclude.

\section{The Birth and Demise of Axion stars}\label{sec:formation}

An axion (a light pseudo scalar resulting from the breaking of a $U(1)_{PQ}$ global symmetry through non-perturbative QCD effects) or axion-like particle (similar to the axion without the restriction that the potential comes from QCD) provides an interesting dark matter candidate.  We will focus throughout on the more generic axion-like particle (ALP), although we will simply refer to them as axions, whose properties we parametrize through the effective Lagrangian
\be\label{eq:axionLagrangian}
\mathcal{L}  = \frac{1}{2}\partial_\mu a\partial^\mu a - \frac{1}{2}m_a^2 a^2 + \frac{\lambda}{4!} a^4 -\frac{\gagamgam}{4}aF_{\mu\nu}\widetilde{F}_{\mu\nu}~,
\ee
with $\widetilde{F}_{\mu\nu} = \frac{1}{2} \epsilon^{\mu\nu\alpha\beta}F_{\alpha\beta}$ and $\gagamgam\equiv\kappa\frac{\alpha}{2\pi f_a}$ where $\kappa$ is a model dependent numerical factor and $f_a$ is the Peccei-Quinn symmetry breaking scale.  Note that the QCD axion also has some model dependence in its coupling to photons, as well as having a coupling to gluons, and for the QCD axion $\lambda \approx 0.2 \frac{m_a^2}{f_a^2}$.  For simplicity, from now on, we take $\lambda = m_a^2/f_a^2$ and consider $\kappa$ a free parameter; we discuss this more in section~\ref{sec:EM_emission}.  In addition to the couplings in (\ref{eq:axionLagrangian}) there is the coupling to gravity, which can result in the formation of bound structures in dense axion environments.

These bound structures, dubbed axion stars, are compact objects that are made of localized and coherently oscillating axion fields; for a thorough review see \cite{Eby:2019ntd} and references therein. They exist through a balance between outward gradient pressure and either self-gravity or attractive axion self-interactions. Below a critical mass, the axion star is stable and referred to as a dilute axion star. If these dilute axion stars find themselves embedded in a gas of unbound axions the unbound axions can condense, increasing the mass of the axion star.  Once the critical mass is reached, the star becomes unstable and contracts to a dense star \cite{Levkov:2016rkk,Eby:2016cnq,Helfer:2016ljl,Chavanis:2017loo,Zhang:2018slz}.  The dense star has a large central axion field value and is short lived.  In the absence of interactions with the standard model this large field value leads to the production of relativistic axions and the star converts a substantial fraction of its mass into these relativistic states \cite{Levkov:2016rkk} which can lead to cosmological constraints \cite{Fox:2023aat}.  However, if the axion couples to the photon the dense star can instead emit copious amounts of photons of frequency $\nu \sim m_a$.  This can occur through a parametric resonance whereby an incoming CMB photon of the same frequency is resonantly amplified.  We will discuss this process in the next section, and determine the necessary size of the axion-photon coupling for the photon emission timescale to be short enough that the dominant energy loss mechanism is to photons, not relativistic axions.  Subsequently we will determine the rate for this process to occur and in which regions of axion parameter space there will be a signal at radio telescopes.

An axion star of mass $\mstar$ and radius $\rstar$ has energy
\be
E_* = -\frac{G_N \mstar^2}{\rstar} + c_1 \frac{\mstar}{2\,m_a^2 \rstar^2} - c_2 \frac{\lambda\mstar^2}{12\,m_a^4 \rstar^3}~.
\ee
The terms in this expression are due to gravitational self energy, gradient pressure, and the internal energy from self interactions and the $c_i$, depend upon the details of the field profile and are found numerically \cite{Ruffini:1969qy,Membrado:1989ke,Visinelli:2017ooc} to be $c_1= 9.9$, $c_2=0.85$.  A good approximation of the numerical solutions is given by a Gaussian field profile \cite{Chavanis:2011zi,Chavanis:2011zm,Chavanis:2016dab}, $a(r)/ f_a = \Theta(r)= \sqrt{2} \Theta_0 {\rm exp}(-r^2/(2R_*^2))$.  

By minimizing the total energy, the mass-radius relation is
\be
\label{eq:mass_radius_relation}
\rstar^\pm=\frac{c_1}{2\,G_N \mstar m_a^2}\left(1\pm \sqrt{1-\frac{c_2}{c_1^2}  \lambda G_N\mstar^2}\right)~.
\ee
The stability of the solution is determined by the sign of $\partial^2E_*/\partial R^2|_{R=R_{\pm}}$. Solutions in the dilute branch ($R=R_{+}$) are stable while those in the critical branch ($R=R_{-}$) are unstable. In the dilute branch, the central field value satisfy $\Theta_0\lesssim f_a/M_{\rm pl}$. The solutions of dilute axion star exist when the quantity below the square root in Eq.~(\ref{eq:mass_radius_relation}) is positive. When that becomes negative, stable solutions of axion field configurations no longer exist. Therefore, axion stars above a critical mass will collapse and enter the dense branch. The critical mass $\mstar^{\mathrm{crit}}$ at which this occurs is
\begin{equation}\label{eq:critical_mass}
	M^{\star}_{\mathrm{crit}}= \frac{10.7\,M_{\rm pl}f_a}{m_a} \approx 1.16\times 10^{-11}M_{\odot}\left(\frac{\rm 10^{-5}eV}{m_a}\right)\left(\frac{f_a}{10^{12}\rm GeV}\right)~.
\end{equation}

We now turn to the rate of formation, and subsequent growth, of the axion stars from a background density of free axions, as in a minihalo.  The relevant timescale is determined by the axion scattering cross section, $\sigma$, the typical density, $n$, and speed, $v$, of the background axions,
\be\label{eq:tau}
\tau \sim \left(f_{\mathrm{BE}} n\sigma v\right)^{-1}~.
\ee
The Bose-enhancement factor, $f_{\mathrm{BE}}= 6\pi^2 n(m_a v)^{-3}$, arises from the large axion phase space density.  See Appendix~\ref{app:halo_appendix} for more discussion of the properties of minihalos.  

In the region of parameter space we will be interested in the scattering cross section is determined by the axion self interactions~\footnote{For larger $f_a$ the gravitational cross section can become important.  The associated timescale for this $\tau_{\mathrm{grav}} \sim m_a v^6/(48\pi^3 G_N^2 n^2$, see \cite{Levkov:2018kau,Eggemeier:2019jsu,Chen:2021oot,Kirkpatrick:2020fwd,Chen:2020cef,Fox:2023aat} for more details.}, $\sigma_{\mathrm{self}}= \lambda^2/(128\pi m_a^2)$ and the resulting timescale is 
\be
\tau_{\rm self}\approx \frac{64 m_a^5 v^2 f_a^4}{3\pi n^2}~.
\ee
This timescale determines the growth rate for axion stars inside minihalos, see Appendix~\ref{app:formation_rate} for how this depends on minihalo mass.  It has recently been shown \cite{Dmitriev:2023ipv} that the mass growth of an axion star is described by a self-similar solution.  Introducing the quantity $\epsilon\approx 3 \overline{M}_\star/M_h$
with the saturation mass defined as $\overline{M}_\star = 3\, G_N^{-1/2} m_a^{-1} \rho_s^{1/6} M_h^{1/3}$.
The ratio of star mass to halo mass $x_{\star}=\mstar(t)/M_h$ was shown to evolve from $x_\star(0)=0$ as
\be
\label{eq:mass_growth}
\frac{(1+x_{\star}^3/\epsilon^2)^3}{(1-x_{\star})^5}\approx \frac{t+0.1\tau}{1.1\tau}~.
\ee
After an initial period of linear growth, and while the axion star mass remains a small fraction of the mass of the halo it forms in, and $t>\tau$, this growth is well described by a power law, $\mstar\sim \overline{M}_\star (t/\tau)^{1/3}$.  At late times when the star has grown to be an appreciable fraction of the halo it is also a power law but of lower power, $\mstar\sim t^{1/9}$.  The total time for an axion star to reach the critical mass can be determined from (\ref{eq:mass_growth}), by solving for the time $t_{\mathrm{crit}}$ when $x_{\star}=\mstar^{\mathrm{crit}}/M_h$. 

Given that the self-similar growth is well approximated by a (changing) power law, and that a simple power law results in closed analytic forms for various results, we choose to present our results for both a simple power law and the more complicated self-similar growth of \cite{Dmitriev:2023ipv}.  If the mass growth is in the power-law region, $\mstar = \overline{M}_\star(t/\tau)^{1/n}$, then the timescale of forming a critical star is $t_{\rm crit}= \tau (M_{\rm crit}/\overline{M}_\star)^n$
and we consider $n$ in the range from 2 to 5. As we will see the qualitative results are very similar for power law and self-similar growth.

Once the mass of the star grows beyond the critical mass the axion self-interactions become important, the star becomes unstable and starts to shrink.  The evolution of the dilute star has been calculated under the assumption of a Gaussian field profile \cite{Chavanis:2016dab}, where it was shown that 
for an axion star just above critical mass the radius initially shrinks slowly $\dot{\rstar}\sim -t$, while at late times it evolves quickly $\dot{\rstar}\sim -(t_{\mathrm{coll}}-t)^{-3/5}$ with total  collapse time given by
\be
\tau_{\mathrm{coll}} \approx 0.49\frac{\lambda}{m_a}\left(\frac{M_{pl}}{m_a}\right)^2\left(\frac{\mstar-\mstar^{\mathrm{crit}}}{\mstar^{\mathrm{crit}}}\right)^{-1/4}~.
\ee
During the period where its radius shrinks and before it emits many axions $\Theta_0  \sim \rstar^{-3/2}$. Once this central field value becomes order-unity $\Theta_0\sim 1$ it enters the dense axion star regime.  The mass-radius relation become $M_{*}\propto \rstar^3$ because the energy density of axion star $\rho\sim\theta^2m_a^2f_a^2$ saturates.  In particular
\cite{Visinelli:2017ooc}, 
\be
\rstar^D\sim 0.6 \left(\frac{\mstar}{m_a^2 f_a^2}\right)^{1/3}\approx \frac{1.3}{m_a}\left(\frac{M_{\rm pl}}{f_a}\right)^{1/3}~,
\ee
where, in the second equality, we have taken the dense axion star mass to be the critical mass discussed in Eq.~\ref{eq:critical_mass} since stable axion stars will collapse to a unstable dense axion star once they accrete to a mass just above the critical value.

Axions in such configurations have  high momenta and they will emit relativistic axions via the self-interaction term, causing the instability of dense axion stars. The life time of dense stars due to the emission of relativistic axions is \cite{Visinelli:2017ooc}: 
\begin{equation}\label{eq:dense_lifetime}
	\tau_{\rm life}=\mathcal{O}\left(\frac{10^3}{m_a}\right)\approx 10^{-7}{\rm s}\left(\frac{10^{-5}\rm eV}{m_a}\right).
\end{equation}
It is worth noting that the collapse time, $\tau_{\mathrm{coll}}$, is much longer than the dense axion star lifetime, $\tau_{\rm life}$, by at least a factor of $(M_{pl}/m_a)^2$. However, the collapse time is still very short compared to cosmological timescales.  So far the evolution of the axion star has ignored all but self interactions.  In the next section we will address what happens during this collapse if there is a sizable coupling of the axion to photons.

\section{Photon Emissions from Axion Star Explosions}\label{sec:EM_emission}

A crucial aspect of the study of axinovae is to determine if they will produce observable photons, which provides new signals in the search of axinovae. It is known that a coherent axion field can experience stimulated axion decay, which is a process whereby the axion decays are further stimulated by outgoing photons, due to Bose enhancements.
This stimulated decay of diffuse axion background has been extensively studied, and can lead to interesting radio line signatures \cite{Arza:2018dcy,Caputo:2018vmy,Arza:2019nta,Arza:2020eik,Sun:2021oqp,Buen-Abad:2021qvj, Sun:2023gic}. Dense coherent axion clumps such as axion stars will further increase the discovery opportunity. 
Recently, there has been considerable progress on stimulated decay in axion clumps or solitons, and the rate estimates of the axion star collapse \cite{Hertzberg:2018zte, Wang:2020zur,Chung-Jukko:2023cow,Di:2023nnb,Levkov:2020txo,Amin:2020vja,Amin:2021tnq,Escudero:2023vgv,Amin:2023imi,Di:2024snm,Di:2024tlz,Maseizik:2024qly,Maseizik:2024uln,Maseizik:2024qya,Di:2024jia}.

There are competing effects that can stop the growth of the photon flux in realistic environments \footnote{We thank Junwu Huang for bringing this to our attention.}.  The presence of charged particles (electrons and positrons) in the region with strong electromagnetic fields can lead to electromagnetic showers and the generation of a plasma, through strong field processes like photon-assisted Schwinger pair production. This, in turn, leads to a mass for the photon which can make the stimulated decay of an axion kinematically blocked~\cite{Arvanitaki:2021qlj,Siemonsen:2022ivj}. In what follows we first estimate the size of the field in a region with only axions and photons, we call this $E_{\mathrm{max,vac}}$, and then estimate the maximum field once the back reaction from charged particles is included, $E_{\mathrm{max,plasma}}$. This will allow us to determine the maximum expected photon signal from dense axion stars.

The coupling between axions and photons (\ref{eq:axionLagrangian}) modifies the equation of motion of photons in the presence of axion fields, leading to the axion-electrodynamics equations \cite{Wilczek:1987mv}.
For now, we will assume we are in a region which is free of matter and ignore the charge and current densities.  We treat the axion field as a fixed background and ignore the back reaction on its dynamics, which remains valid until axions dominantly decay to photons.  The profile of the axion star is as described in the previous section and we consider the axion field to be coherent over the whole star,
\be
a = f_a \Theta(r) \cos \left(m_a t + \delta\right)~,
\ee
where $\Theta(r)$ represents the dimensionless axion field value, which can reach order unity in dense axion stars.
Furthermore, the profile of the star is slowly varying with $|\partial_r \Theta| \ll m_a, |\vecb{k}|$, with $\vecb{k}$ the wavevector of the light. In the dense branch of axion stars, relativistic modes with momentum $3m_a$ start to appear but do not dominate the axion profile \cite{Visinelli:2017ooc}. Also, dense axion stars are not as compact as black holes as long as $f_a\ll M_{\rm pl}$, which is the parameter space we focus on in this work that can lead to interesting observational signatures.
Therefore, we expect the approximations to hold even for dense axion stars in the parameters of interest. 

These dense objects can lead to interesting lensing effects \cite{Prabhu:2020pzm,Hertzberg:2018zte,Levkov:2020txo,McDonald:2020why,Croon:2020ouk,Yin:2024xov}, dynamical heating of stars \cite{Graham:2023unf,Graham:2024hah} or radio signals from encounters with neutron stars   \cite{Bai:2021nrs,Witte:2022cjj,Kouvaris:2022guf,Maseizik:2024qly}.  
We make the simplifying assumption that the light is normally incident on the axion star and we take the direction of propagation to be along the $z$-axis so that the electric and magnetic fields lie in the $x-y$ plane and can be described in terms of left and right circularly polarized fields
$
\vecb{B}_{\pm} = B_\pm(t) e^{i k z} \bm{\epsilon}_\pm ~,
$
and similarly for $\vecb{E}$.  Using the assumption that the axion star profile is smooth we solve the equations of motion in the adiabatic approximation \ie~we solve for constant axion field and then allow the axion to slowly vary within that solution.  The equation of motion for the electromagnetic field (focussing on the B-field) becomes
\be
\ddot{B}_\pm + k^2 B_\pm = \mp \gagamgam m_a f_a k\, \Theta(z) \sin \left(m_a t + \delta\right) B_\pm~.
\label{eq:BpmEOM}
\ee
Defining $\eta=(m_a t+\delta)/2$ we see that \eref{eq:BpmEOM} takes on the form of the Mathieu equation,
\be
B^{''}_\pm + \left[\left(\frac{2k}{m_a}\right)^2  \pm \frac{4\gagamgam f_a k}{m_a}  \Theta(z) \sin 2\eta \right] B_\pm = 0~,
\label{eq:MathieuEq}
\ee
with primes denoting derivatives with respect to $\eta$.  The Mathieu equation is known to exhibit the phenomenon of parametric resonance.  For $4\gagamgam f_a k\, \Theta(z)\ll m_a$ parametric resonance can occur only for photon frequencies close to half the axion mass. Both the ingoing and reflected photons receive amplification, 
$ 
B_\pm(t)\sim \frac{1}{\sqrt2} \left(e^{i k z} + e^{-i k z}\right)\,e^{\mu t}
$,
with the growth rate $\mu$, also called the \emph{Floquet} exponent, given by \cite{Hertzberg:2018zte}
\be
\mu = \frac{1}{2} \sqrt{\Big(\gagamgam f_a \Theta(z) k \Big)^2-\frac{m_a^2}{4}\left(1-\left(\frac{2k}{m_a}\right)^2\right)^2}~.
\ee
The Floquet exponent is real, and thus parametric resonance occurs, only for photon momenta close to $m_a/2$.  For such photons the maximum growth rate is 
\be
\overline{\mu} = \frac{\gagamgam m_a f_a \Theta\, }{4}\equiv \mu_0 \Theta~,
\ee
which is linear in the coupling $\gagamgam$. The width of the resonance band is thus $\Delta k \sim m_a f_a g_{a\gamma\gamma}/4$.
Photons that take a path of length $L$ through the axion star, we ignore possible focussing effects and assume the photon's path is unaffected by the axion background, will have their amplitude boosted by $\exp (\int_0^L\overline{\mu}(\eta) d\eta)$.  The typical path length of photons through the star is $L\sim R_*$.     

Modeling the axion profile as a Gaussian within the axion star, \ie\ $\Theta(r) = \sqrt{2}\Theta_0 {\rm exp}(-r^2/(2R_*^2))$, with the normalisation given by $\Theta_0\sim (\frac{M_{*}}{\pi^{3/2}R_*^3\,m_a^2f_a^2})^{1/2}$ the enhancement in the amplitude of the electromagnetic field as a photon pass through the axion star scales as $\propto e^{2\sqrt{\pi}\mu_0 \Theta_0 R_*}$.
For a dilute critical mass axion star where $\Theta_0\sim f_a/M_{\mathrm{pl}}$ and $R_*\sim M_{\mathrm{pl}}/(m_a f_a)$ this enhancement is small, $\mu_0 \Theta_0 R_*\sim \kappa \alpha$.  However for (short lived) dense axion stars $\Theta_0\sim 1$, resulting in a potentially large enhancement $\mu_0 \Theta_0 R_*^D\sim \kappa \alpha\left(\frac{M_{\rm pl}}{f_a}\right)^{1/3}$

This exponential enhancement of electric field strength may be seeded by either the incoming CMB photons are spontaneous decay of an axion, but the enhancement can be so large that the final field is insensitive to the initial field value.  As an estimate we consider the initial field value to be that coming from CMB photons, with frequency in the resonance band \ie $|k-m_a/2|<\bar{\mu}$.  The incoming field strength is $E_{\mathrm{cmb}}=(T_{\mathrm{cmb}} m_a^2 \bar{\mu}/4\pi^2)^{1/2}$. This can result in a very high photon field,
\be\label{eq:Eamp}
E_{\mathrm{max,vac}} 
\approx
\sqrt{\kappa}\left(\frac{m_a}{10^{-6}\eV}\right)^{3/2}\exp\left(1.4\kappa \left(\frac{10^{10}\GeV}{f_a}\right)^{1/3}\right)\times 10^{-33}\GeV^2~.
\ee
The maximum possible E-field is exponentially sensitive to the axion-photon coupling, $\kappa$, and in models with enhanced couplings this final E-field can be very large.

As discussed earlier, there is another effect that can potentially cap the production of photons in the axion star, related to Schwinger pair production \cite{PhysRev.82.664}. Although the field strength never gets large enough for the vacuum Schwinger effect, which occurs when $e E\sim m_e^2$, there are additional catalysed-Schwinger effects that can occur at lower electric field.  In particular, in the presence of high frequency photons, with energy $\omega_\gamma$, there is the photon-assisted Schwinger effect \cite{Schutzhold:2008pz,Dumlu:2010ua,Bashmakov:2013iwa} which occurs when
\be\label{eq:assistedSchwinger}
E\sim \frac{2m_e^3}{e \omega_\gamma}~,
\ee
for $\omega_\gamma>m_e$ this is a lower field value than the vacuum Schwinger effect.

The interior of the axion star is likely to contain free charges, in which case the growing photon field grows will accelerate them, giving them Lorentz factor $\gamma$.  If they are accelerated perpendicular to their motion they in turn will emit synchrotron radiation of frequency $\omega_\gamma \sim \gamma^3 \dot{v}_{\perp}$.  If these frequencies are high enough they can lead to pair production and these electrons/positrons in turn will be accelerated. There is a run away process and the interior of the star is rapidly filled with an electron-positron plasma.  The resulting plasma mass for the photon blocks the decay of axions.  To estimate the maximum allowed E-field we need to determine $\omega_\gamma$.

We expect that the E-field inside the axion star is coherent over length scales of $d_E$ and thus that the radius of curvature, $d_c$, of electrons as they pass from one coherent region to another is also $d_c\sim d_E$.  Thus, their Lorentz factor is $\gamma \sim e E d_E/m_e$ and $\dot{v}_{\perp}\sim 1/d_c\sim 1/d_E$.  The bound from (\ref{eq:assistedSchwinger}) means the maximum possible E-field is
\be
E_{\mathrm{max}} \sim e^{-1}\sqrt{\frac{m_e^3}{d_E}}~.
\ee 
Naively we expect the coherence length to be $\mathcal{O}(m_a^{-1})$ which places a constraint on the maximal possible E-field of
\be\label{eq:Emax}
E_{\mathrm{max,plasma}} \lsim \sqrt{\frac{m_a}{10^{-6}\eV}}\times 10^{-12}\GeV^2~.
\ee
Alternatively, if the E-field is coherent across the whole star and the radius of curvature of its motion is similar then $E_{\mathrm{max}} \lsim \left(\frac{m_a}{10^{-6}\eV}\right)^{1/2} \left(\frac{f_a}{10^{10}\GeV}\right)^{1/6}\times 10^{-14}\GeV^2$. We take the maximal possible E-field to be the smaller of (\ref{eq:Eamp}) and (\ref{eq:Emax}), \ie\ $E_{\mathrm{max}}=\min(E_{\mathrm{max,vac}}, E_{\mathrm{max,plasma}})$.  The corresponding energy emitted in photons, with frequency $m_a/2$ is $ E_{\mathrm{max}}^2 R_*^3$.  This energy is released in a time of order $1000/m_a$, which is roughly the light crossing time and the lifetime of dense axion stars. 

In typical axion models the $\kappa$ parameter is $\mathcal{O}(1)$.  However, there are various classes of models with enhanced axion-photon couplings.  For instance, if there is a heavy vectorlike fermion carrying PQ charge $Q_{PQ}$ and electromagnetic charge $Q$ then $\kappa = 2 Q_{PQ} Q^2$, although the requirement that there be no Landau poles below the Planck scale limits $\kappa \ltap 100$ \cite{Agrawal:2017cmd}.  The clockwork mechanism can be used to generate a photo-philic axion \cite{Farina:2016tgd}.  Kinetic mixing between multiple axions \cite{Agrawal:2017cmd} or multiple $U(1)$'s \cite{Daido:2018dmu} can increase the axion photon coupling.  Alternatively, one could suppress the axion self-interactions \cite{DiLuzio:2021pxd} which raises  the critical axion star mass and consequently the radius of the dense axion star, leading to greater enhancement of the electromagnetic field as a photon passes through the dense star.

Over the region of axion mass parameter space that present radio telescopes have access to, and thus we are most interested in, the axion mass lies in the range $10^{-7}\eV \ltap m_a \ltap 10^{-5}\eV$.  Thus, for moderately enhanced couplings, $\kappa\sim 20-30$, the parametric resonance effect is large and the electric field is limited by the assisted Schwinger effect \ie\ ($E_{\rm max,vac}>E_{\rm max,plasma}$).   In our result plot, \cref{fig:ma_fa_fluence}, we will assume that $\kappa$ is large enough that $E_{\rm max}=E_{\rm max,plasma}$ throughout the plotted parameter space.  

\section{Brightness of Visible Axinovae and Observational Constraints} \label{sec:observation}

In the previous sections we have demonstrated that, with a sufficiently large axion-photon coupling and in certain environments, axion stars can form, grow, and then have an axinova event where they emit a large number of photons.  The rate for this signal depends upon the distribution of minihalos in the cosmos, since the growth rate of stars depends upon the density and speed of axions in the core of minihalos.  In what follows we will first determine the minihalo mass spectrum, using the Press-Schechter technique applied to the white noise power spectrum expected for post-inflation PQ breaking.  Then with this spectrum in hand we can determine the rate of axinovae averaged over the halo mass distribution.  In determining the axinovae rate we will consider both a simple power law growth rate and the self-similar growth rate \cite{Dmitriev:2023ipv} for the axion star mass.  Since all axion stars go critical at the same mass, and thus emit the same number of radio photons they are a standard candle \cite{Di:2024tlz} and the signal at a radio telescope will be dominated by the closest axion star.  The typical distance to an axinova can be determined once the rate of axinovae is known.  This distance will determine the fluence of radio photons at the telescope.  We compare this fluence to the noise level in a typical radio telescope, to finally determine which axion parameters produce a sufficiently large signal (we take this to be a signal-to-noise ratio larger than $10$) to be observed/excluded.  Rather than attempt to predict the signal at a particular radio facility we use parameters, like collection area, noise temperature, observing time, frequency binning \etc, of an SKA-like radio telescope when surveying the blank sky \ie\ when not looking at a known radio source.

For the breaking of PQ symmetry after inflation the initial spectrum of axion density perturbations at very small scales follows a white noise power law \ie\ $\langle\delta^2\rangle \sim A_0 (k/k_0)^3$.  After matter-radiation equality these perturbations grow and will form axion minihalos.  As minihalos merge this growth will continue up until late times, $z\sim 20$, at which time the minihalos will be absorbed by larger CDM structures, which have grown from the (approximately) scale-invariant density perturbations of $\Lambda$CDM.  After being absorbed their growth essentially freezes.  To determine the growth from matter-radiation equality to $z\sim 20$ we follow the Press-Schechter approach \cite{1974ApJ...187..425P} with a white-noise spectrum to determine the halo mass function $df_0(M_h,z)/dM_h$.  Similarly, the CDM halo mass function, $df_{\mathrm{col}}/dM_h$, can be calculated using Press-Schechter but with a scale invariant spectrum at large scales.  The final, late time, halo mass function is then calculated as
\begin{equation}
	\frac{{\rm d}f}{{\rm d}M}(z) =\int_{z_{\rm eq}}^{z} {\rm d}z_{\rm i} \frac{{\rm d}f_{\rm col}^{\rm CDM}(z_{\rm i})}{{\rm d}z_{\rm i}}\frac{{\rm d}f_{0 }}{{\rm d}M}(z_{\rm i})+ (1-f_{\rm col}^{\rm CDM}(z)) \frac{df_0}{dM}(z)~.
\end{equation}
This is discussed in more detail in Appendix~\ref{app:minihalo}.  

When determining the signal rate we take into account all these effects, but the evolution of minihalos is dominated by what happens at $z>20$ and thus depends mostly on $df_0(M_h,z)/dM_h$.  In this range of $z$ the growth function also takes on a simple form $D_{\mathrm{grow}}\sim (1+z)^{-2}$.  This results in a relatively simple form for the halo mass function, which is zero for $M< M_0$ and for $M>M_0$ is
\be
\frac{df_0}{d \log M_h}(M_h,z) = \frac{\delta_c}{3\pi}\sqrt{\frac{2M_h}{A_0 M_0}}\left(\frac{1+z}{1+z_{\mathrm{eq}}}\right)\, \exp\left[ -\frac{2\delta_c^2}{9\pi A_0}\left(\frac{1+z}{1+z_{\mathrm{eq}}}\right)^2 \frac{M}{M_0}\right]~,
\ee
where $M_0$ is the the initial minicluster mass, determined by the particle horizon when the axion starts to oscillate $M_0\sim k_{\mathrm{osc}}^3 \rho_0$, which in turn can be determined from $m_a,f_a$, see Appendix~\ref{app:formation_rate} for details.
Notice that $\frac{df_0}{d \log M_h}$ only depends on halo mass through the ratio $M_h/M_0$.
We will present analytic results below using this halo mass function, although we emphasize the final (numerical) results use the full form outlined in Appendix~\ref{app:minihalo}.

The rate of axinova per unit volume is equal to the number density of halos of a given mass divided by the time it takes for those halos to nova,
\be\label{eq:decay_rate}
\frac{d\Gamma}{dM_h}(M_h,z)=\Theta(t_H-t_{\rm crit}) \frac{\rho(z)}{M_h}\frac{df(M_h,z)}{dM_h}\frac{1}{t_{\mathrm{crit}}(M_h,z, m_a, f_a)}~,
\ee
here $\rho(z) = \rho_0 (1+z)^3$ is the cosmological dark matter density at redshift $z$.  Here $t_{\rm crit}$ is the timescale of forming an axion star at the critical mass. 
The total rate can be found by integrating over all halo masses that contain a critical star, \ie\ from $\max (M_0, \mstar^{\mathrm{crit}})$. This integral is dominated by the lightest halos.

The distance, $D_\star$, to the closest axinova within an observing period $\Delta t_{\mathrm{obs}}$ can be found from the total rate $D_\star\sim (\Gamma(z=0)\Delta t_{\mathrm{obs}})^{-1/3}$.  
For instance, if $M_0>\mstar^{\mathrm{crit}}$ and for simplicity we take the growth of the star mass to scale as a simple power $\mstar\sim t^{1/2}$ then $t_{\rm crit}=\tau(M_{\rm crit}/\overline{M}_\star)^2 $ the distance to the closest axinova is
\be
D \sim \left(\frac{f_{\rm sky} \Delta t_{\mathrm{obs}} }{10^{-3} \mathrm{day} } \right)^{-1/3} \left(\frac{m_a}{10^{-5}\eV}\right)^{1/3} \left(\frac{f_a}{10^{10}\GeV}\right)^{4/3}\times 1.3\,\mathrm{kpc}~.
\ee
Note that we have replaced the cosmic dark matter density with the local dark matter density for consistency since the distance is very small for axion masses at radio frequencies.

The radio signal from an axion star consists of a sizeable amount of energy ($\sim E_{\rm max,plasma}^2 R_{*}^{3}$) released in a very short period of time ($\tau_{\rm life}\sim 10^3/m_a\sim 1 \rm \mu s$) in the form of photons peaked around frequency ($\nu\sim m_a/2$) with a spread of $\delta\nu_s\sim m_a f_a g_{a\gamma\gamma}/4\sim \rm 10\,MHz$, but the location on the sky is unknown.  Due to the short duration and small spectral spread the underlying flux is large.  For instance, the fluence (spectral flux density integrated over the duration of the burst) from this source, assuming the axion-photon coupling is large enough to obtain a saturated E field (\ref{eq:Emax}), is given by 
\begin{equation}\label{eq:fluence_AS}
	F_{\rm an}\sim \frac{(4\pi/3) (E_{\rm max}^2/2) R_{*}^{3}}{4\pi D^2 \delta\nu_s}\sim  \frac{0.1}{\kappa\alpha} \left(\frac{\Delta t_{\mathrm{obs}}}{\mathrm{day}}\right)^{2/3}\left(\frac{f_{\rm sky}}{10^{-3}}\right)^{2/3} \left(\frac{m_a}{10^{-5}\rm eV}\right)^{-2/3}\left(\frac{f_a}{10^{10}\rm GeV}\right)^{-8/3}\, \rm Jy\cdot ms,
\end{equation}
where $\rm Jy= 10^{-26}W\cdot m^{-2}\cdot Hz^{-1}$ is the Jansky unit.

However, although radio telescopes can have very fine frequency resolution and $\mu s$ timing resolution when looking at a known radio source a broad sky survey does not collect data with such fine resolution.  Typically the radio flux is integrated over periods of order a second, $\Delta t\sim 1\mathrm{sec}$ and the frequencies are binned with $\Delta \nu\sim 100 \mathrm{kHz}-1 \mathrm{MHz}$ resolution.  This time and frequency binning is considerably broader than the intrinsic duration and spectral spread and so dilutes the signal.  One must compare the energy collected from the axion star during a single bin in time and frequency space to the expected background in the same bin.  The background is often described in terms of the noise equivalent temperature $T_{\mathrm{sys}}$, which is the temperature a resistor with the same power output.  The noise level over the frequency range ($\Delta\nu$) and duration of the observation ($\Delta t$), to which the signal needs to be compared, is $T_{\mathrm{sys}}/\sqrt{\Delta\nu\Delta t}$.  With typical noise temperatures of $50\mathrm{K}$ the background fluence is $\sim 10^{-13}\mathrm{Jy\ ms\ Hz}^{-1}$.  Thus, 
the expected signal to noise ratio in a radio telescope of collecting area $A$ is
\be\label{eq:StoN}
\frac{S}{N}=F_{\rm an} A \delta\nu_s \frac{\sqrt{\Delta\nu\Delta t}}{T_{\mathrm{sys}}} 
~.
\ee

Taking benchmark values for current radio telescopes the signal-to-noise ratio can be numerically expressed as 
\be\label{eq:SNR}
\frac{S}{N}=6.5 \times 10^{3}
\left(\frac{A}{1000 \mathrm{m}^2}\right) \left(\frac{m_a}{10^{-5}\eV}\right)^{-8/3} \left(\frac{f_a}{10^{10}\GeV}\right)^{-11/3}
\left(\frac{f_{\mathrm{sky}} \Delta t_{\mathrm{obs}}}{10^{-3}\mathrm{day}}\right)^{2/3} \left(\frac{50\mathrm{K}}{T_{\mathrm{sys}}}\right) 
\left(\frac{\Delta t}{\mathrm{1\,ms}}\right)^{1/2} \left(\frac{\Delta \nu}{100 \mathrm{MHz}}\right)^{1/2} 
~.
\ee

\begin{figure}[h!] 
	\centering
	\includegraphics[width=0.85\textwidth]{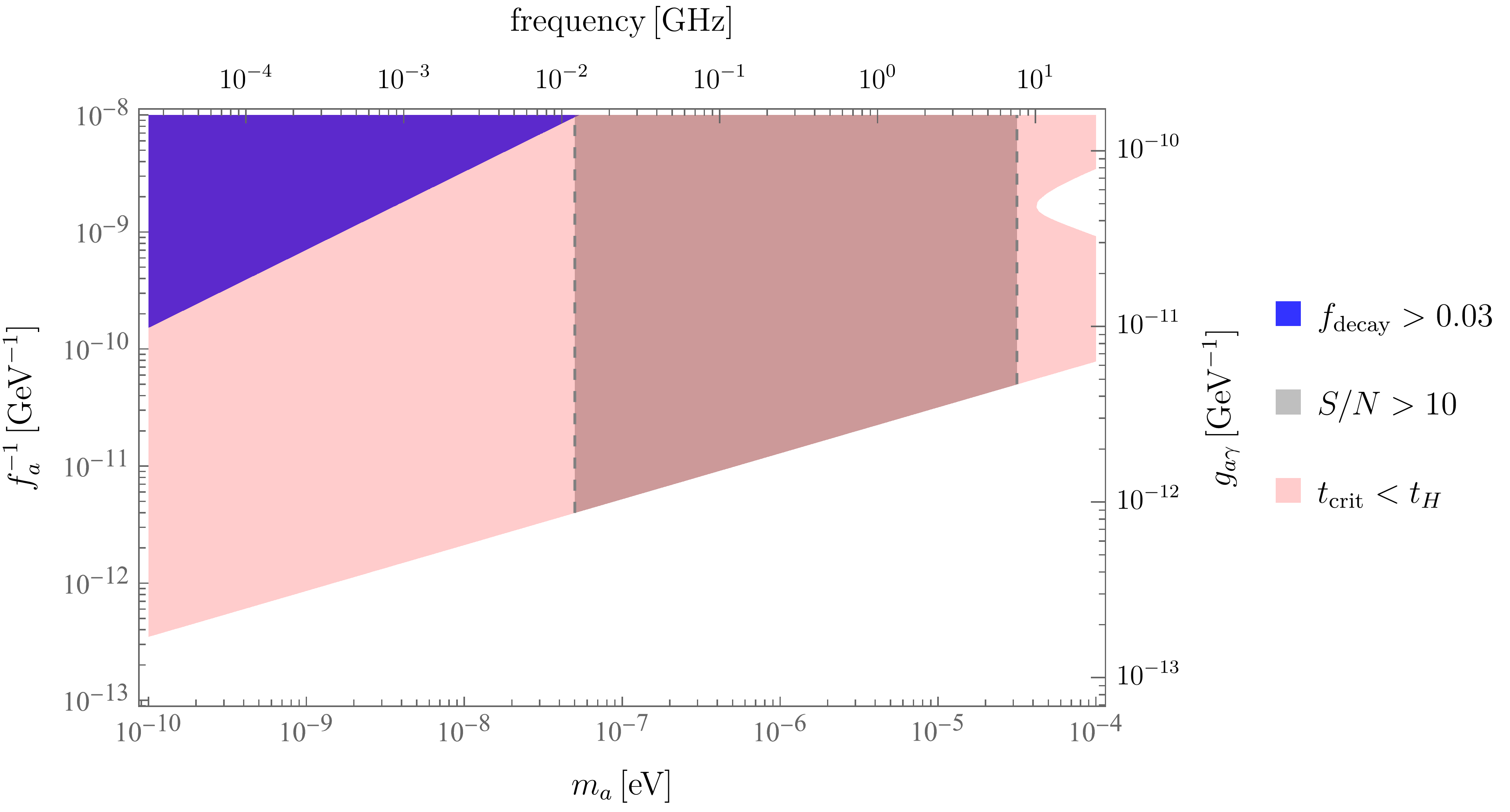} 
	\caption{Axion parameters that lead to axinovae in the current Universe for power law growth of axion star mass, $M_*\propto t^2$. The blue region is excluded by cosmological observations due to the large axinova rate.  The pink region cannot be excluded by cosmology but can lead to an observable rate of axinova. The gray region represents the axion parameters that lead to sizable radio signals with a fluence above the detection threshold of radio telescopes, corresponding to a signal-to-noise ratio ($S/N$) greater than 10.}
	\label{fig:ma_fa_fluence}
\end{figure}

Radio telescopes have sensitivity over a range of radio frequencies, with the lower edge ($\nu \gtap 10^{-2}\,\mathrm{GHz}$) set by the opacity of the Earth's atmosphere and the upper edge ($\nu\ltap 15\,\mathrm{GHz}$) determined by the observing window of, for example, SKA\cite{SKA:2018ckk}.  In Fig.~\ref{fig:ma_fa_fluence} we denote this approximate region of observability by gray shading and with dashed lines indicating the frequency range of radio observations.

\begin{figure}[h!] 
	\centering
	\includegraphics[width=0.85\textwidth]{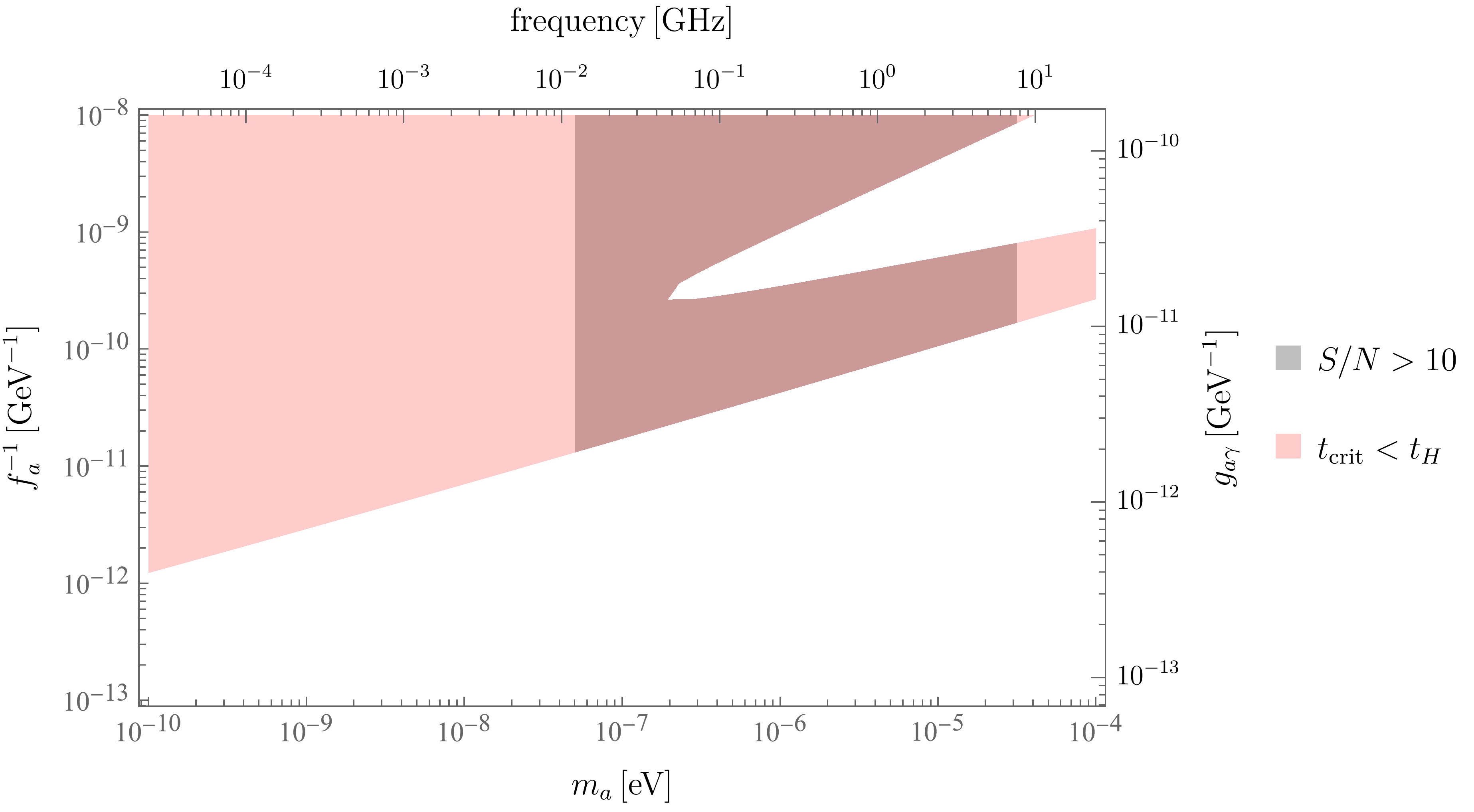} 
	\caption{Axion parameters that lead to axinovae in the current Universe using the self-similar growth in Ref.~\cite{Dmitriev:2023ipv}. The pink region represents the axion parameter space that leads to axion star collapse and the gray region corresponds to radio emissions with $S/N>10$. Note that the result is very similar to that in Fig.~\ref{fig:ma_fa_fluence} even though we have used a different formula for axion star growth.
	}
	\label{fig:ma_fa_Levkov}
\end{figure}

In Figs.~\ref{fig:ma_fa_fluence} and \ref{fig:ma_fa_Levkov} we present the region of parameter space that leads to a significant radio signal.  As discussed above, over the range of frequencies for which a radio telescope is sensitive, the size of the electric field that can be generated in the axion star is limited by the assisted Schwinger effect, and this requires a modest enhancement of the axion photon coupling.  The necessary enhancement is slightly $m_a, f_a$ dependent, and we assume $\kappa$ is large enough throughout the mass range shown in the plot.  The pink shaded region is where the lighter halos in the halo mass function can form a critical axion star within the present age of the universe.  In Fig.~\ref{fig:ma_fa_fluence} we follow \cite{Fox:2023aat} and take the growth in mass of the axion star to follow a simple power law, with $n=2$, in Fig.~\ref{fig:ma_fa_Levkov} we use the self-similar solution for growth (\ref{eq:mass_growth}).

For smaller $f_a$ the formation and subsequent explosion rate can be high enough to cause changes to our cosmological history, as discussed in Ref.~\cite{Fox:2023aat}.  This region is shaded blue, and we present a closed-form for the decay fraction, under the assumption that  $n=2$, in the appendix (\ref{eq:dfdz}).  Note that this region relies upon the decay rate of axion stars, which does not depend on the electromagnetic interactions of axions, which only determines the energy fraction carried by photons during axion star collapse.  There is a slight difference between the recurrent axinovae constraint presented in \cite{Fox:2023aat} and the blue region in Fig.~\ref{fig:ma_fa_fluence}, which is related to assumptions about the dependence of the axion mass on temperature.  In the present work we have fixed the initial axion minicluster mass by determining the oscillation time from the relic abundance as shown in Eq.~(\ref{eq:minicluster_mass}) assuming the axion oscillates at its zero temperature mass, whereas in \cite{Fox:2023aat} we fixed the temperature when oscillation began and altered form of $m_a(T)$ to achieve the correct relic abundance.  The pink region below the blue region has critical axion star production, but not at sufficient rate to cause large scale structure to be altered.  There is a possible electromagnetic signal and we denote in gray the region where a radio telescope with receiving area of $1000\, \mathrm{m}^2$ observing $0.1\%$ of the sky for 24 hours would see a signal-to-noise of $10$, see Eq.~\ref{eq:SNR}.  If the photons emitted from axions in the pink region are not in the radio band, radio observation constraints do not apply~\footnote{For lower frequencies, it is also required that the plasma frequency is smaller than $m_a/2$ to not kinematically block photon emissions from axion star collapse, which requires axion with masses $m_a >2\times 10^{-14}$ eV to explode in the current Universe \cite{Berlin:2022hmt}.}, which form the left and right vertical boundary of the gray region. Comparing the results of Fig.~\ref{fig:ma_fa_fluence} and Fig.~\ref{fig:ma_fa_Levkov} we see that simple power law growth and self-similar growth give qualitatively similar results for the signal at a radio telescope. We have not presented the cosmological signal in Fig.~\ref{fig:ma_fa_Levkov} for the self-similar growth because it needs to be reparametrized to account for the recurrent axinova rate, which is not the focus in this work. 
The region of axion parameter space with a potentially accessible radio signal, shown in \cref{fig:ma_fa_fluence,fig:ma_fa_Levkov}, is not presently excluded by other search techniques \cite{AxionLimits}.  There are several proposals for future experiments that will have reach into the region shown in \cref{fig:ma_fa_fluence,fig:ma_fa_Levkov}.

\section{Conclusion}
\label{sec:conclusion}

In this work, we addressed the question of whether axion stars can form cosmologically and whether axion star explosions (axinovae) can emit photons resulting in observable signals. The cosmological scenario of axions considered in this work is the post-inflationary scenario where the Peccei-Quinn symmetry breaking occurs after inflation, we leave the pre-inflationary scenario for future studies. In the post-inflationary scenario, the axion field is randomized in different horizon patches before it acquires its mass, resulting in large isocurvature perturbations roughly at the horizon size of initial axion oscillation. Such density perturbation will lead to the formation of dense substructures called axion miniclusters, which subsequently merge and form larger objects called axion minihalos. These dense substructures are ideal environments for axion star formation due to their large density and small virial velocity. 

The calculation of the formation rate of axion stars and the event rate of their explosions extends our previous work \cite{Fox:2023aat} to include the full mass spectrum of axion minihalos whereas before we focussed only on halos masses at the peak of the mass spectrum.  
We systematically study the stimulated decay of axions within dense axion stars, taking into account the effects of parametric resonance which enhances the photon emission as well as the photon-assisted Schwinger effect which hinders the emission.  Despite the fact that these dense axion stars are the short-lived end state of an axion star and will decay away after a few thousand axion oscillations, we show that in models with moderately enhanced axion-photon couplings there can be sufficient photon emission to allow detection.  Existing radio telescopes observing the blank sky for relatively short periods of time have the potential to see one of these transient objects. Over much of the accessible parameter space, which is not presently excluded by other search techniques, the rate for seeing these events is sufficiently high that the parameter space would likely be quickly excluded by a dedicated search, or a signal would be quickly found.  We advocate for such a search.

\bigskip

\noindent\textbf{Note added:} While this work was being finalised a related analysis appeared \cite{Alexander:2025flo} using mergers of axion stars in multi-axion models to produce radio signals from axinovae.

\section*{Acknowledgements}
The authors thank Jae Hyeok Chang, Raymond Co, Junwu Huang, Kiyoshi Masui, Albert Stebbins, Yitian Sun, Wei Xue for helpful discussions.
Fermilab is managed by FermiForward Discovery Group, LLC, acting under Contract No.89243024CSC000002 with the U.S. Department of Energy, Office of Science, Office of High Energy Physics.  
Part of this research was performed at the Aspen Center for Physics, which is supported by National Science Foundation grant PHY-1607611.
N.W. is supported by NSF under award PHY-2210498, by the BSF under grant 2018140, and by the Simons Foundation.
\appendix

\section{The Mass Function of Axion Minihalos and the Decay Rate of Axion Stars}\label{app:minihalo}
Enhanced substructures from white-noise perturbations can greatly accelerate the gravitational condensation of axions because of the high density of those structures. Therefore it is crucial to determine the mass function of those axion substructures to calculate the population of axion stars.

A simple model of collapsed halos by Press $\&$ Schechter \cite{1974ApJ...187..425P} determines the probability that a halo forms as
\begin{equation}\label{eq:mass_fraction_function}
	\frac{df_0}{d M_h} = \frac{1}{M_h} \sqrt{\frac{\nu}{2\pi}}{\rm exp}(-\nu/2)\left|\frac{d\,{\rm ln}\nu}{d\,{\rm ln}M_h}\right|~,
\end{equation}
where $\nu$ is defined as $\nu \equiv \frac{\delta_c^2}{\sigma^2(M_h)D(z)^2}$,
with $\delta_c=1.686$ the overdensity threshold for spherical collapse and $D(z)$ is the growth function. In a matter dominated cosmology the growth function takes the simple form $D(z)=1+\frac{3}{2}(1+z_{\mathrm{eq}})/(1+z)$.
The variance of the smoothed density field is denoted $\sigma^2(M)$ and is defined as
\begin{equation}
	\sigma^2(M)= \int \frac{dk}{k}\frac{k^3P(k)}{2\pi^2}\vert W(kR)\vert^2~.
\end{equation}
There are many choices for the smoothing procedure.  We use a simple top-hat filter function whose Fourier transform is $W(x) = (3/x^3)[{\rm sin}(x)-x{\rm cos}(x)]$ and for which $M=(4\pi/3)R^3\rho_0$.   For a white-noise dominated power spectrum $\frac{k^3P(k)}{2\pi^2}=A_{\rm osc}(k/k_{\rm osc})^3$ and the variance is
\begin{equation}
	\sigma(M)=\sqrt{\frac{3\pi A_{\rm osc}}{2}\frac{M_0}{M}}~, 
\end{equation}
with $M_0= \frac{4\pi}{3}k_{\rm osc}^{-3}\rho_{0}$.  The mass distribution of axions (\ie\ $M_h df_0/dM_h$) is peaked at mass scale
\be
M_{\rm peak} = \frac{3\pi A_{\mathrm{osc}}}{2\delta_c^2}
D(z)^2 M_0~,
\ee
which corresponds to $\nu = 1$.  The minihalo mass function (\ref{eq:mass_fraction_function}) predicted by the Press-Schechter approach agrees well with N-body simulations with an initial white-noise power spectrum \cite{Xiao:2021nkb}.

The above mass function applies for time before the formation of large scale structures that are seeded from primordial density perturbation, which occurs at $z\sim 20$.  
At later times the axion minihalos can be absorbed by massive CDM structures, freezing the subsequent mergers of axion minihalos. 
The final axion minihalo mass function, including infall minihalos that have been absorbed by standard CDM halos can be written as
\begin{equation}
	\frac{{\rm d}f}{{\rm d}M}(z) =\int_{z_{\rm eq}}^{z} {\rm d}z_{\rm i} \frac{{\rm d}f_{\rm col}^{\rm CDM}(z_{\rm i})}{{\rm d}z_{\rm i}}\frac{{\rm d}f_{0 }}{{\rm d}M}(z_{\rm i})+ (1-f_{\rm col}^{\rm CDM}(z)) \frac{df_0}{dM}(z)~.
\end{equation}
where $f_{\rm col}^{\rm CDM}(z)$ is the collapse fraction of massive CDM halos formed from adiabatic fluctuations. 
The probability of infall at a particular time slice $z=z_{\rm i}$ is proportional to ${\rm d}f_{\rm col}^{\rm CDM}(z_{\rm i})/{{\rm d}z_{\rm i}}$, where we have assumed the axion minihalos will be the building blocks of massive CDM halos during their collapse. This assumption is valid as long as there is a mass hierarchy between CDM halos and axion minihalos.
Again applying the Press-Schechter formalism for standard CDM halos, their collapse fraction can be written as 
\begin{equation}
	f_{\rm col}^{\rm CDM}(z)={\rm erfc}\left(\frac{\delta_{\rm c}}{\sqrt{2}\,\sigma_{\rm CDM}(M_{\rm min})\,D(z)}\right),
	\label{eqn:fcol}
\end{equation}
where $\sigma^{2}_{\rm CDM}(M)$ is the variance of the adiabatic fluctuations and $M_{\rm min}$ is the minimal mass of CDM halos that can absorb axion minihalos. With the above expressions, we can compute the full evolution of axion minihalo mass function all the way from matter radiation equality to the current time.

\section{Halo properties}\label{app:halo_appendix}

The density profile and mass function of axion minihalos are needed if we want to compute the condensation time scale and determine the fate of axion stars. Therefore, it is crucial to know the evolution of the mass and size of axion miniclusters or minihalos over cosmic time. Axion miniclusters or minihalos are highly concentrated substructures with  a Navarro-Frenk-White (NFW)\cite{Navarro_1996} density profile, which is parametrized by a scale density $\rho_s$ and a scale radius, $r_s$.  The density profile is
\be\label{eq:NFW}
\rho(r)=\frac{\rho_s}{\frac{r}{r_s}\left(1+\frac{r}{r_s}\right)^2}~.
\ee
The circular speed at the scale radius, which we will take as indicative of the speed of particles forming the star is
\be\label{eq:circspeed}
v_s^2 = 4\pi G_N \rho_s r_s^2\left(\log 4-1\right)~.
\ee
It is conventional, assuming spherical collapse, to introduce the concentration, defined as $c=r_{200}/r_s$, where $r_{200}$ is the radius at which the average halo density is $200$ times the background dark matter density \ie
\be
\frac{M_h}{\frac{4\pi}{3}r_{200}^3} = 200(1+z)^3\rho_{DM,0}~.
\ee
This then allows $r_s$ to be determined in terms of halo mass and concentration and $\rho_s$ to be determined in terms of concentration,
\be\label{eq:rhos}
\rho_s = \frac{\frac{200 c^3}{3}}{\log(1+c)-\frac{c}{1+c}}\left(1+z\right)^3\rho_0~.
\ee
Thus, the timescale for axion star formation can be determined once the mass and concentration of the minihalo is known.  From numerical simulations \cite{Dai:2019lud,Xiao:2021nkb} is has been shown that the earliest halos form with  $c \approx 4$ and subsequently those that grow into the largest halos keep this concentration while lighter halos that form later have higher concentration, due to the redshift of the dark matter density.  This behaviour is well fit by
\be
c(M_h,z) = \frac{1.4\times 10^4}{(1+z)\sqrt{\frac{M_h}{A_0 M_0}}}~.
\ee

\section{Axion Star Formation Time and Axion Minihalo Formation Mass}\label{app:formation_rate}

The timescale for an axion star that grows inside a minihalo depends upon the density and speeds of axions in the halo.  The distribution of axions inside NFW halos is discussed in Appendix~\ref{app:halo_appendix}.  The axion star is likely to form at the center of the halo, but to be conservative we estimate the condensation time at the scale radius where the density is lower and the speed higher, both of which raise the estimate of the evolution timescale $\tau$.  For a scattering cross section dominated by self interactions the timescale, relative to the Hubble time is,
\begin{equation}
	\tau_{\rm self}H_0\approx \frac{64 H_0 m_a^3v^2f_a^4}{3\pi\rho^2}\approx  \left(\frac{m_a}{ \mu \rm eV}\right)^3\left(\frac{f_a}{10^9\,\rm GeV}\right)^4\left(\frac{M_0}{10^{-12}M_{\odot}}\right)^{2/3}\left(\frac{M_h/M_0}{10}\right)^{19/6}~,
\end{equation}
where $M_0$ is a characteristic mass scale for initial axion miniclusters, and $M_h$ is the axion minihalo mass.  We have suppressed a weak dependence on $z$.

The characteristic mass, $M_0$, which appears throughout these discussions, is determined by when the axion first starts to oscillate which, for the vacuum misalignment mechanism, occurs when $m_{\rm osc}\equiv m_a(T_{\mathrm{osc}}) = 3H(T_{\mathrm{osc}})/2$.  Thus,
\be
M_0 = \frac{4\pi}{3}\left(a(T_{\mathrm{osc}})H_{\mathrm{osc}}\right)^{-3}  \bar{\rho}_{a,0}~.
\ee
The exact oscillation temperature is model dependent and in some sense the characteristic mass of axion minihalos, $M_0$, remains a free parameter in the theory. 
However, $M_0$ can be determined from the oscillation time of axions, which is related to its relic abundance.  The relic abundance is given by
\begin{equation}
	\Omega_a =\frac{m_a m_{\rm osc}f_a^2\langle\theta^2\rangle}{2\rho_{\rm crit}}\frac{g_*(T_0)T_0^3}{g_*(T_{\rm osc})T_{\rm osc}^3},
\end{equation}
where $T_0$ is the current CMB temperature and $m_{\rm osc}$ is the mass of the axion when it starts to oscillate, at $T=T_{\rm osc}$. For simplicity, we assume $m_{\rm osc}=m_a$, which is a good approximation for an axion mass which rapidly changes from $0$ to $m_a$ during the phase transition of the dark confinement sector.  Thus, as $m_a, f_a$ vary the requirement of the correct relic abundance determines $T_{\rm osc}$ which in turn determines $M_0$.
We only consider the relic abundance from vacuum misalignment and take $\langle\theta^2\rangle=\pi^2/3$.  While axions from axion string decays might also contribute to the relic abundance these are not expected to dominate \cite{OHare:2021zrq}. Therefore, the initial minicluster mass can be determined by the ALP mass as
\begin{equation}\label{eq:minicluster_mass}
	M_0 \approx 1.5\times 10^{-12}M_{\odot} \left(\frac{m_a}{10^{-6}\rm \, eV}\right)^{-2}\left(\frac{f_a}{10^{11}\rm \, GeV}\right)^{-2}\left(\frac{g_{*}(T_{\rm osc})}{100}\right)^{-1/2}~.
\end{equation}

Once we have the minicluster mass, as well as the mass function, the fraction of axion stars per redshift that has become critical and collapsed can be calculated. We can obtain an analytic formula assuming a power-law mass growth $M\propto t^{1/2}$ \cite{Fox:2023aat}:
\begin{equation}
	\label{eq:dfdz}
	\begin{split}
		\frac{d f_{\rm decay}}{dz}&\sim
			\int_{M_0}^{\infty} dM \frac{df}{dM}\,	\frac{M_{\rm pl}^3 \overline{\rho}_{\rm col}(M)^2}{M  f_a^5 m_a^4} \frac{1.5\times 10^6 \,\pi^{2/3}\kappa}{(1+z)^{5/2}H_0}   \left[1+6.8\pi^{4/3}\left(\frac{f_a}{M^{1/3}\overline{\rho}_{\rm col}(M)^{1/6}}\right)^4\right]
		\boldsymbol{H} \left(M-	M^{\star}_{\mathrm{crit}}\right)~,
	\end{split}
\end{equation}
where $\boldsymbol{H}(x)$ is the Heaviside unit step function, $\kappa=0.1$ is the efficiency of mass conversion to radiation during axinovae, $df/dM$ is the mass function of axion minihalos discussed in Appendix~\ref{app:minihalo}, $z_c$ is the redshift axions start to collapse and form axion miniclusters, which we take as $z_c=z_{\rm eq}$, and $\overline{\rho}_{\rm col}(M)\approx \, \rho_{\rm eq}(M/M_0)^{-3/2}$ is the scale density of axion minihalos for given halo mass $M$ with $\rho_{\rm eq}$ being the matter density at matter radiation equality. $M^{\star}_{\mathrm{crit}}$ is the critical mass of axion stars calculated in Eq.~\ref{eq:critical_mass}. Calculating the above expression at the current time, we can obtain the axion parameter space that will lead to axion star explosions.

\begin{figure}[h!] 
	\centering
	\includegraphics[width=0.85\textwidth]{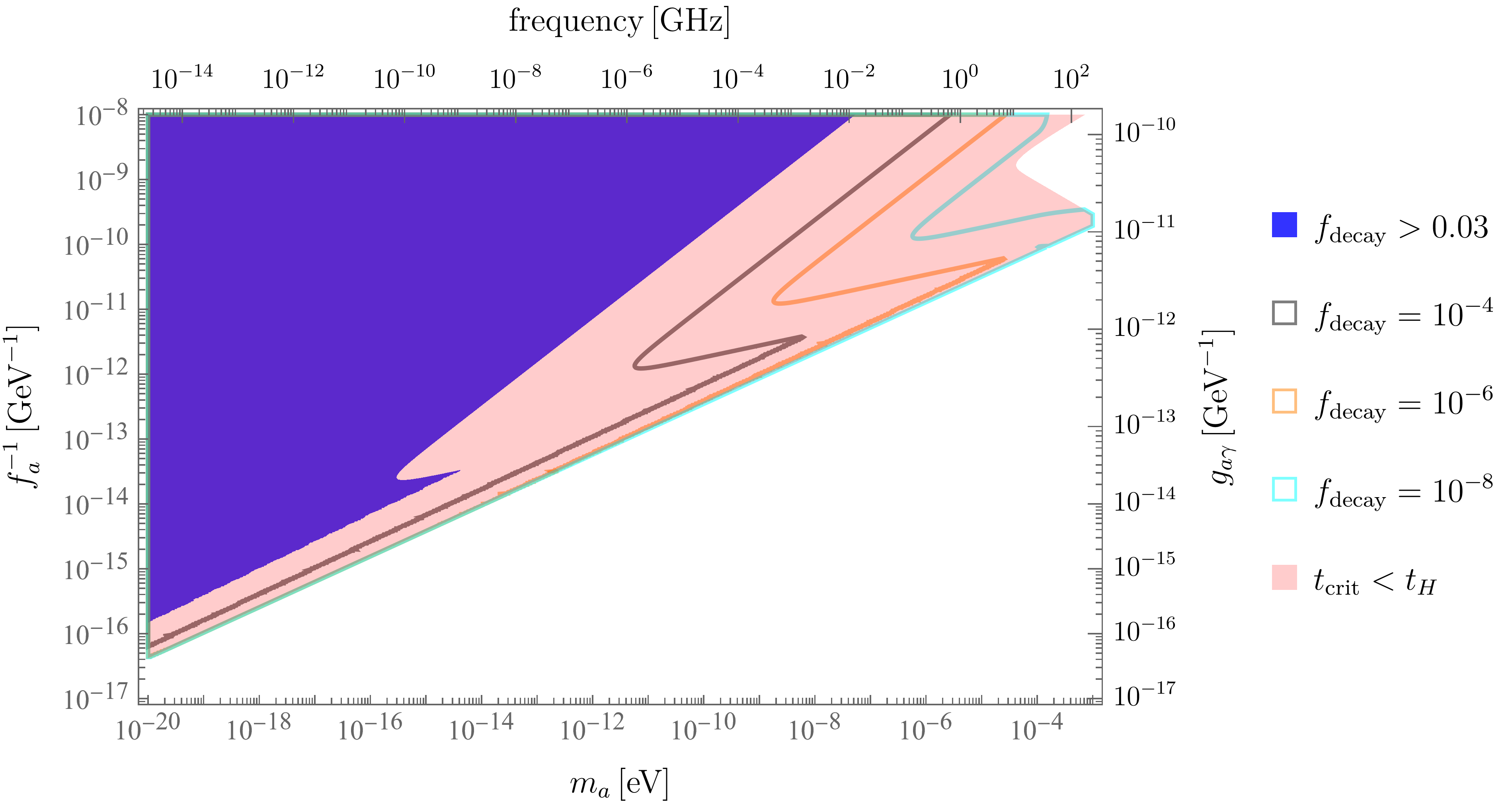} 
	\caption{Axion parameters that lead to different decay fraction of axion dark matter from axinovae assuming power law growth. We added a few more contours that indicate different decay fraction. The blue region is excluded by cosmological observations. This plot presents a wider range of parameters compared to Fig.~\ref{fig:ma_fa_fluence} and more contours on different decay fraction but otherwise showing the same physical results.
	}
	\label{fig:ma_fa_dfdz}
\end{figure}

Now we can compute the decay rate of axion dark matter induced by axinovae as a function of axion parameters $m_a,f_a$ using Eq.~\ref{eq:dfdz}, which gives the result in Fig.~\ref{fig:ma_fa_dfdz}. A lower $f_a$, indicating a stronger self-coupling and a smaller critical mass, will result in higher axinova rate. However, a larger $f_a$ will correspond to an earlier oscillation time when we fix relic abundance, which gives a smaller axion minicluster mass and subsequently an enhanced axion star formation rate.  
One could also consider exotic cosmological scenarios \cite{Nelson:2018via,Visinelli:2018wza,Blinov:2019jqc} or other production mechanism \cite{Co:2019jts,Chang:2019tvx,Eroncel:2022efc,Eroncel:2022vjg,Redi:2022llj,Harigaya:2022pjd}, which will lead to a different result on the axion minicluster mass. 

\bibliographystyle{apsrev4-2}
\bibliography{AxionStar}
\end{document}